\documentclass[]{llncs}

\usepackage[]{textcomp}
\usepackage[]{microtype}
\usepackage[utf8]{inputenc}
\usepackage[T1]{fontenc}
\usepackage[]{upgreek}
\usepackage[]{mathtools}
\usepackage[]{amssymb}

\usepackage{hyperref}

\let\oldampersand\&

\renewcommand*\&{{\itshape\oldampersand}}

\title{Balancing lists: a proof pearl\thanks{This research has received funding from the European Research Council under the FP7 grant agreement 278673, Project MemCAD}}

\titlerunning{Balancing lists}

\author{Guyslain Naves\inst{1} and Arnaud Spiwack\inst{2}}

\institute{Aix Marseille Université, CNRS, LIF UMR 7279, 13288, Marseille, France\\
\email{guyslain.naves@lif.univ-mrs.fr}\and Inria Paris-Rocquencourt\\\textsc{Ens}, Paris, France\\
\email{arnaud@spiwack.net}}

\begin{document}
  \maketitle
  \begin{abstract}
    Starting with an algorithm to turn lists into full trees which uses non-obvious invariants and partial functions, we progressively encode the invariants in the types of the data, removing most of the burden of a correctness proof.
    \par
    The invariants are encoded using non-uniform inductive types which parallel numerical representations in a style advertised by Okasaki, and a small amount of dependent types.
  \end{abstract}
  \section{Introduction}
  \par
  Starting with a list \textsf{lst}, we want to turn it into a binary tree \textsf{tr} of the following form (in Ocaml):
  \par
  \begin{displaymath}
    \parbox{0.9\linewidth}{\textbf{\textsf{type}}~${\upalpha}$~\textsf{tree}~\texttt{\symbol{61}}\\
    \hphantom{ }\hphantom{ }\hspace*{0.1em}\rule[-0.6ex]{1.sp}{1.\baselineskip}\hspace*{0.2em}~\textsf{Node}~\textsf{of}~${\upalpha}$~\textsf{tree}~\texttt{\symbol{42}}~${\upalpha}$~\texttt{\symbol{42}}~${\upalpha}$~\textsf{tree}\\
    \hphantom{ }\hphantom{ }\hspace*{0.1em}\rule[-0.6ex]{1.sp}{1.\baselineskip}\hspace*{0.2em}~\textsf{Leaf}}
  \end{displaymath}
  \par
  With the constraints that \textsf{lst} must be the infix traversal of \textsf{tr} and that \textsf{tr} must be \emph{full}, in the sense that every level except the last are required to be completely filled. Such a function turns, in particular, sorted lists into balanced binary search trees.
  \par
  There are a number of folklore algorithms to achieve this result in linear time. Here we consider one of these algorithms, presented in Section~\ref{latex_lib_label_1}, which repeatedly pairs up trees of height $h$ in a list to form a list of trees of height $h+1$. Our interest in this algorithm sprouts from the fact that its correctness is not obvious; in particular the invariants are complex: the main loop operates on a list of length $2^k-1$ whose elements are alternately of two distinct forms.
  \par
  In Sections~\ref{latex_lib_label_2} and~\ref{latex_lib_label_3} we show refinements of the algorithm where the invariants are pushed into the types, leading to a complete and short proof of correctness in Coq.
  \section{A balancing algorithm}\label{latex_lib_label_1}
  \par
  We start by giving a first, simple, implementation of the balancing algorithm. The heart of the algorithm relies on using an alternating list of length ${2}^{k}-1$, where odd-position elements are trees and even-position elements are labels, of type ${\upalpha}$ (indices starting from 1).  A full tree of height $k$ can be decomposed into the first $k-1$ levels, containing ${2}^{k-1} -1$ internal nodes, and the $k$th level, which contains both nodes and leaves. Thus, the ${2}^{k-1} - 1$ labels in the alternating list will be used to label the internal nodes in the $k-1$ first levels of the balanced tree, while the ${2}^{k-1}$ trees, all of height at most one at first, will constitute the level $k$.
  \par
  Though we could encode labels as trees of height one in the alternating list, we rather use an appropriate type for the sake of readability:
  \par
  \begin{displaymath}
    \parbox{0.9\linewidth}{\textbf{\textsf{type}}~${\upalpha}$~\textsf{tree\texttt{\symbol{95}}or\texttt{\symbol{95}}elt}~\texttt{\symbol{61}}\\
    \hspace*{0.1em}\rule[-0.6ex]{1.sp}{1.\baselineskip}\hspace*{0.2em}~\textsf{Elt}~\textsf{of}~${\upalpha}$\\
    \hspace*{0.1em}\rule[-0.6ex]{1.sp}{1.\baselineskip}\hspace*{0.2em}~\textsf{Tree}~\textsf{of}~${\upalpha}$~\textsf{tree}}
  \end{displaymath}
  \par
  We decompose the problem into two parts: computing an alternating list of length ${2}^{k} - 1$ from an arbitrary list of labels, and then transforming this alternating list into a balanced tree. We first show how to solve the second part: turning an alternating list into a full tree.
  \par
  Given an alternating list \textsf{lst}, by pairing the trees in \textsf{lst} using only one traversal of the list, we obtain an alternating list with exactly half as many trees. Each pairing requires two trees and one label used as a root. In order to build a list that is alternated, we also need a second label, that is kept as a single element. This explains why we consider at each step the four first elements of the list.
  \par
  A single traversal, encoded by \textsf{pass}~\texttt{\symbol{58}}~${\upalpha}$~\textsf{tree\texttt{\symbol{95}}or\texttt{\symbol{95}}elt}~\textsf{list}~${\rightarrow}$~${\upalpha}$~\textsf{tree\texttt{\symbol{95}}or\texttt{\symbol{95}}elt\texttt{\symbol{95}}list}, reduces an alternating list of length ${2}^{k} - 1 \geqslant  3$ to an alternating list of length ${2}^{k-1} -1$. By iterating this process using \textsf{loop}~\texttt{\symbol{58}}~${\upalpha}$~\textsf{tree\texttt{\symbol{95}}or\texttt{\symbol{95}}elt}~\textsf{list}~${\rightarrow}$~${\upalpha}$~\textsf{tree}, we reduce the original list to a list of length one, whose one element is a balanced tree \textsf{t} such that the infix traversal of \textsf{t} is the initial list. 
  \par
  \begin{displaymath}
    \parbox{0.9\linewidth}{\textbf{\textsf{let}}~\textsf{join}~\textsf{left}~\textsf{node}~\textsf{right}~\texttt{\symbol{61}}~\textsf{Tree}~\texttt{\symbol{40}}\textsf{Node}~\texttt{\symbol{40}}\textsf{left}\texttt{\symbol{44}}~\textsf{node}\texttt{\symbol{44}}~\textsf{right}\texttt{\symbol{41}\symbol{41}}\\[0.5\baselineskip]
    \textbf{\textsf{let}}~\textbf{\textsf{rec}}~\textsf{pass}~\texttt{\symbol{61}}~\textbf{\textsf{function}}\\
    \hphantom{ }\hphantom{ }\hspace*{0.1em}\rule[-0.6ex]{1.sp}{1.\baselineskip}\hspace*{0.2em}~\textsf{Tree}~\textsf{left}~\texttt{:\hspace*{-0.25em}:}~\textsf{Elt}~\textsf{root}~\texttt{:\hspace*{-0.25em}:}~\textsf{Tree}~\textsf{right}~\texttt{:\hspace*{-0.25em}:}~\textsf{Elt}~\textsf{e}~\texttt{:\hspace*{-0.25em}:}~\textsf{others}~${\rightarrow}$\\
    \hphantom{ }\hphantom{ }\hphantom{ }\hphantom{ }\hphantom{ }\hphantom{ }\textsf{join}~\textsf{left}~\textsf{root}~\textsf{right}~\texttt{:\hspace*{-0.25em}:}~\textsf{Elt}~\textsf{e}~\texttt{:\hspace*{-0.25em}:}~\textsf{pass}~\textsf{others}\\
    \hphantom{ }\hphantom{ }\hspace*{0.1em}\rule[-0.6ex]{1.sp}{1.\baselineskip}\hspace*{0.2em}~\texttt{\symbol{91}}\textsf{Tree}~\textsf{left}\texttt{\symbol{59}}~\textsf{Elt}~\textsf{root}\texttt{\symbol{59}}~\textsf{Tree}~\textsf{right}\texttt{\symbol{93}}~${\rightarrow}$~\texttt{\symbol{91}}\textsf{join}~\textsf{left}~\textsf{root}~\textsf{right}\texttt{\symbol{93}}\\
    \hphantom{ }\hphantom{ }\hspace*{0.1em}\rule[-0.6ex]{1.sp}{1.\baselineskip}\hspace*{0.2em}~\texttt{\symbol{95}}~${\rightarrow}$~\textsf{assert}~\textsf{false}\\[0.5\baselineskip]
    \textbf{\textsf{let}}~\textbf{\textsf{rec}}~\textsf{loop}~\texttt{\symbol{61}}~\textbf{\textsf{function}}\\
    \hphantom{ }\hphantom{ }\hspace*{0.1em}\rule[-0.6ex]{1.sp}{1.\baselineskip}\hspace*{0.2em}~\texttt{\symbol{91}\symbol{93}}~${\rightarrow}$~\textsf{Leaf}\\
    \hphantom{ }\hphantom{ }\hspace*{0.1em}\rule[-0.6ex]{1.sp}{1.\baselineskip}\hspace*{0.2em}~\texttt{\symbol{91}}\textsf{Tree}~\textsf{t}\texttt{\symbol{93}}~${\rightarrow}$~\textsf{t}\\
    \hphantom{ }\hphantom{ }\hspace*{0.1em}\rule[-0.6ex]{1.sp}{1.\baselineskip}\hspace*{0.2em}~\textsf{list}~${\rightarrow}$~\textsf{loop}~\texttt{\symbol{40}}\textsf{pass}~\textsf{list}\texttt{\symbol{41}}}
  \end{displaymath}
  \par
  Notice how the invariant that alternating lists have length $2^k-1$ is maintained: this is because, for $k \geqslant  2$, we have $2^k - 1 = 4 \times  ({2}^{k-2} - 1) + 3$, hence we obtain an alternating list of length $2 \times  ({2}^{k-2} - 1) + 1 = {2}^{k-1} - 1$.
  \par
  It remains to show how to transform a list of labels of length $n$ into an alternating list of trees and labels. Each of the original trees has height zero or one: they are leaves or contain only one label. Because we want a list of length precisely $2^k -1$, for $k = 1 + \left\lfloor {\log }_{2} n\right\rfloor $, it means we need $2^k - 1 - n$ leaves. This quantity is computed as the variable \textsf{missing}. The function \textsf{pad} computes the alternating list by creating as many leaves as needed, alternating them with elements, and once enough leaves are created, promotes all the odd-position labels into trees.
  \par
  \begin{displaymath}
    \parbox{0.9\linewidth}{\textbf{\textsf{let}}~\textsf{complete}~\textsf{list}~\texttt{\symbol{61}}\\
    \hphantom{ }\hphantom{ }\textbf{\textsf{let}}~\textsf{n}~\texttt{\symbol{61}}~\textsf{List}\texttt{\symbol{46}}\textsf{length}~\textsf{list}~\textbf{\textsf{in}}\\
    \hphantom{ }\hphantom{ }\textbf{\textsf{let}}~\textbf{\textsf{rec}}~\textsf{pow2}~\textsf{i}~\texttt{\symbol{61}}~\textsf{if}~\textsf{i}~\texttt{\symbol{60}\symbol{61}}~\textsf{n}~\textsf{then}~\textsf{pow2}~\texttt{\symbol{40}2\symbol{42}}\textsf{i}\texttt{\symbol{41}}~\textsf{else}~\textsf{i}~\textbf{\textsf{in}}~\\
    \hphantom{ }\hphantom{ }\textbf{\textsf{let}}~\textsf{missing}~\texttt{\symbol{61}}~\texttt{\symbol{40}}\textsf{pow2}~\texttt{1\symbol{41}}~\texttt{\symbol{45}}~\textsf{n}~\texttt{\symbol{45}}~\texttt{1}~\textbf{\textsf{in}}\\
    \hphantom{ }\hphantom{ }\textbf{\textsf{let}}~\textbf{\textsf{rec}}~\textsf{pad}~\textsf{missing}~\texttt{\symbol{61}}~\textbf{\textsf{function}}\\
    \hphantom{ }\hphantom{ }\hphantom{ }\hphantom{ }\hspace*{0.1em}\rule[-0.6ex]{1.sp}{1.\baselineskip}\hspace*{0.2em}~\textsf{head}\texttt{:\hspace*{-0.25em}:}\textsf{tail}~\textsf{when}~\textsf{missing}~\texttt{\symbol{60}\symbol{62}}~\texttt{0}~${\rightarrow}$\\
    \hphantom{ }\hphantom{ }\hphantom{ }\hphantom{ }\hphantom{ }\hphantom{ }\hphantom{ }\hphantom{ }\textsf{Tree}~\textsf{Leaf}~\texttt{:\hspace*{-0.25em}:}~\textsf{Elt}~\textsf{head}~\texttt{:\hspace*{-0.25em}:}~\textsf{pad}~\texttt{\symbol{40}}\textsf{missing}~\texttt{\symbol{45}}~\texttt{1\symbol{41}}~\textsf{tail}\\
    \hphantom{ }\hphantom{ }\hphantom{ }\hphantom{ }\hspace*{0.1em}\rule[-0.6ex]{1.sp}{1.\baselineskip}\hspace*{0.2em}~\textsf{odd}\texttt{:\hspace*{-0.25em}:}\textsf{even}\texttt{:\hspace*{-0.25em}:}\textsf{others}~${\rightarrow}$~\textsf{join}~\textsf{Leaf}~\textsf{odd}~\textsf{Leaf}~\texttt{:\hspace*{-0.25em}:}~\textsf{Elt}~\textsf{even}~\texttt{:\hspace*{-0.25em}:}~\textsf{pad}~\texttt{0}~\textsf{others}\\
    \hphantom{ }\hphantom{ }\hphantom{ }\hphantom{ }\hspace*{0.1em}\rule[-0.6ex]{1.sp}{1.\baselineskip}\hspace*{0.2em}~\texttt{\symbol{91}}\textsf{single}\texttt{\symbol{93}}~${\rightarrow}$~\texttt{\symbol{91}}\textsf{join}~\textsf{Leaf}~\textsf{single}~\textsf{Leaf}\texttt{\symbol{93}}\\
    \hphantom{ }\hphantom{ }\hphantom{ }\hphantom{ }\hspace*{0.1em}\rule[-0.6ex]{1.sp}{1.\baselineskip}\hspace*{0.2em}~\texttt{\symbol{91}\symbol{93}}~${\rightarrow}$~\texttt{\symbol{91}\symbol{93}}\\
    \hphantom{ }\hphantom{ }\textbf{\textsf{in}}\\
    \hphantom{ }\hphantom{ }\textsf{pad}~\textsf{missing}~\textsf{list}}
  \end{displaymath}
  \par
  The balancing algorithm \textsf{balance}\texttt{\symbol{58}}~${\upalpha}$~\textsf{list}~${\rightarrow}$~${\upalpha}$~\textsf{tree}~ is thus given by the composition of \textsf{loop} with \textsf{complete}:
  \begin{displaymath}
    \parbox{0.9\linewidth}{\textbf{\textsf{let}}~\textsf{balance}~\textsf{list}~\texttt{\symbol{61}}~\textsf{loop}~\texttt{\symbol{40}}\textsf{complete}~\textsf{list}\texttt{\symbol{41}}}
  \end{displaymath}
  \par
  As for the complexity of this algorithm, notice that \textsf{pass} and \textsf{complete} are both clearly in linear-time in the length of the lists on which they work, while \textsf{loop} recurses on lists whose length are halved at each recursive step. Hence \textsf{balance} is a linear-time algorithm.
  \section{Removing partial functions}\label{latex_lib_label_2}
  \par
  The \textsf{loop} function of Section~\ref{latex_lib_label_1} relies on the invariant that the \textsf{list} argument has length $2^k-1$. Additionally, all the odd-position values must be of the form \textsf{Tree}~\textsf{t}, whereas all the even-position values must be of the form \textsf{Elt}~\textsf{x}. If either of these invariants is broken, we would run into the \textsf{assert}~\textsf{false} of \textsf{pass}.
  \par
  It is not immediately apparent that these properties hold. If it does not take a tremendous effort to convince oneself that the \textsf{balance} function of Section~\ref{latex_lib_label_1} is indeed correct, a direct mechanically checked proof would not be very practical.
  \par
  \subsection{Length invariants}
  \par
  Our goal in this section is to avoid resorting to \textsf{assert}~\textsf{false}. In addition to making sure that \textsf{balance} indeed terminates with a value, it will make it considerably simpler to implement the balancing algorithm in Coq in Section~\ref{latex_lib_label_3}.
  To achieve this goal, it is necessary to have more precise types.
  Let us focus first on the length invariants: we will need to define a type which contains exactly the non-empty lists of length $2^k-1$.
  \par
  A data structure which holds $2^k-1$ elements brings complete binary trees to mind. Even if it is possible -- though not necessary convenient -- to represent complete binary trees in Ocaml, they are not the appropriate structure. First, because complete binary trees are full trees and are, hence, unlikely to serve as a useful intermediate data structure to build a full tree. Second because there is a simpler -- albeit more exotic -- alternative.
  \par
  Indeed, lists can be seen as decorated unary numbers: there is an element at each successor. Different kinds of lists can be obtained, more or less systematically, by varying the numerical representation. This idea goes back to Guibas \& al. in~\cite{Guibas1977} and a fairly thorough exploration can be found in Okasaki~\cite[Chapters 9\&10]{Okasaki1999}. In the simplest cases, the analogous list structure corresponds to a structurally recursive exponentiation algorithm. For regular lists, a list of size $n$ whose elements have type $a$ can be recursively defined with the following equations:
  \begin{displaymath}
    \left\{ 
    \begin{array}{lll}
      {a}^{0} & = & 1\\
      {a}^{n+1} & = & a\times a^n\\
    \end{array}
    \right. 
  \end{displaymath}
  Replacing unary numbers with binary numbers, we obtain the binary exponentiation algorithm:
  \begin{displaymath}
    \left\{ 
    \begin{array}{lll}
      {a}^{2^0-1} & = & 1\\
      {a}^{2n} & = & {(a^2)}^{n}\\
      {a}^{2n+1} & = & a\times {(a^2)}^{n}\\
    \end{array}
    \right. 
  \end{displaymath}
  Okasaki~\cite[Chapter 10]{Okasaki1999} uses a non-uniform inductive type to encode the latter exponentiation algorithm into a type of lists he calls \emph{binary lists}. We are only interested in lists of length $2^k-1$, that is a length written only with the digit $1$ in binary representation. So following Okasaki, but skipping the second equation above (which corresponds to the digit $0$) we define the following non-uniform inductive type, which we call \emph{power lists}:
  \begin{displaymath}
    \parbox{0.9\linewidth}{\textbf{\textsf{module}}~\textsf{PowerList}~\texttt{\symbol{61}}~\textbf{\textsf{struct}}\\[0.5\baselineskip]
    \hphantom{ }\hphantom{ }\textbf{\textsf{type}}~${\upalpha}$~\textsf{t}~\texttt{\symbol{61}}\\
    \hphantom{ }\hphantom{ }\hphantom{ }\hphantom{ }\hspace*{0.1em}\rule[-0.6ex]{1.sp}{1.\baselineskip}\hspace*{0.2em}~\textsf{Zero}\\
    \hphantom{ }\hphantom{ }\hphantom{ }\hphantom{ }\hspace*{0.1em}\rule[-0.6ex]{1.sp}{1.\baselineskip}\hspace*{0.2em}~\textsf{TwicePlusOne}~\textsf{of}~${\upalpha}$~\texttt{\symbol{42}}~\texttt{\symbol{40}}${\upalpha}$\texttt{\symbol{42}}${\upalpha}$\texttt{\symbol{41}}~\textsf{t}\\[0.5\baselineskip]
    \textbf{\textsf{end}}}
  \end{displaymath}
  \par
  This type actually appears in Okasaki~\cite[Chapter 10]{Okasaki1999} as an introduction to non-uniform binary lists. Relatedly, Okasaki~\cite{Okasaki1999a} leverages a tail-recursive binary exponentiation algorithm to define a type capturing precisely square matrices; on the other hand, Myers~\cite{Myers1983} introduced a flavour of list based on \emph{skew binary numbers} which are not easily captured as exponentiation.
  \par
  Although the power lists may look like some sort of trees, it is not a very accurate depiction. The easiest way to picture how power lists works is to see \textsf{TwicePlusOne} as a fancy \texttt{\symbol{40}}\texttt{:\hspace*{-0.25em}:}\texttt{\symbol{41}}, then the lists with, respectively, $1$, $3$, $7$, and $15$ elements are as follows:
  \begin{itemize}
    \item \texttt{\symbol{91}1\symbol{93}}
    \item \texttt{\symbol{91}1\symbol{59}\symbol{40}2\symbol{44}3\symbol{41}\symbol{93}}
    \item \texttt{\symbol{91}1\symbol{59}\symbol{40}2\symbol{44}3\symbol{41}\symbol{59}\symbol{40}\symbol{40}4\symbol{44}5\symbol{41}\symbol{44}\symbol{40}6\symbol{44}7\symbol{41}\symbol{41}\symbol{93}}
    \item \texttt{\symbol{91}1\symbol{59}\symbol{40}2\symbol{44}3\symbol{41}\symbol{59}\symbol{40}\symbol{40}4\symbol{44}5\symbol{41}\symbol{44}\symbol{40}6\symbol{44}7\symbol{41}\symbol{41}\symbol{59}\symbol{40}\symbol{40}\symbol{40}8\symbol{44}9\symbol{41}\symbol{44}\symbol{40}10\symbol{44}11\symbol{41}\symbol{41}\symbol{44}\symbol{40}\symbol{40}12\symbol{44}13\symbol{41}\symbol{44}\symbol{40}14\symbol{44}15\symbol{41}\symbol{41}\symbol{41}\symbol{93}}
  \end{itemize}
  Elements appear in order, like in a regular list, but they are packed twice as tightly after each \textsf{TwicePlusOne}.
  \par
  Just like with regular lists, there is a \emph{map} function for power lists. Due to the non-uniformity it is a little trickier\footnote{The type annotation on \textsf{PowerList}\texttt{\symbol{46}}\textsf{map} informs Ocaml that \textsf{map} is a non-uniform recursive function. Without the type annotation, Ocaml simply assumes that \textsf{map} is uniformly recursive and fails to typecheck since \textsf{f} and \textsf{f}\texttt{\symbol{39}} have different types.} than the regular list map: in the recursive steps, the argument function \textsf{f} needs to process two consecutive elements at a time.
  \begin{displaymath}
    \parbox{0.9\linewidth}{\textbf{\textsf{module}}~\textsf{PowerList}~\texttt{\symbol{61}}~\textbf{\textsf{struct}}\\
    \hphantom{ }\hphantom{ }\hspace{1.em}\raisebox{-1.ex}[0.75\baselineskip][0.75\baselineskip]{{\vdots}}\\
    \hphantom{ }\hphantom{ }\textbf{\textsf{let}}~\textbf{\textsf{rec}}~\textsf{map}~\texttt{\symbol{58}}~${\upalpha}$~${\upbeta}$\texttt{\symbol{46}}~\texttt{\symbol{40}}${\upalpha}$${\rightarrow}$${\upbeta}$\texttt{\symbol{41}}~${\rightarrow}$~${\upalpha}$~\textsf{t}~${\rightarrow}$~${\upbeta}$~\textsf{t}~\texttt{\symbol{61}}~\textbf{\textsf{fun}}~\textsf{f}~${\rightarrow}$~\textbf{\textsf{function}}\\
    \hphantom{ }\hphantom{ }\hphantom{ }\hphantom{ }\hspace*{0.1em}\rule[-0.6ex]{1.sp}{1.\baselineskip}\hspace*{0.2em}~\textsf{Zero}~${\rightarrow}$~\textsf{Zero}\\
    \hphantom{ }\hphantom{ }\hphantom{ }\hphantom{ }\hspace*{0.1em}\rule[-0.6ex]{1.sp}{1.\baselineskip}\hspace*{0.2em}~\textsf{TwicePlusOne}~\texttt{\symbol{40}}\textsf{elt}\texttt{\symbol{44}}\textsf{lst}\texttt{\symbol{41}}~${\rightarrow}$\\
    \hphantom{ }\hphantom{ }\hphantom{ }\hphantom{ }\hphantom{ }\hphantom{ }\hphantom{ }\hphantom{ }\textbf{\textsf{let}}~\textsf{f}\texttt{\symbol{39}}~\texttt{\symbol{40}}\textsf{x}\texttt{\symbol{44}}\textsf{y}\texttt{\symbol{41}}~\texttt{\symbol{61}}~\textsf{f}~\textsf{x}~\texttt{\symbol{44}}~\textsf{f}~\textsf{y}~\textbf{\textsf{in}}\\
    \hphantom{ }\hphantom{ }\hphantom{ }\hphantom{ }\hphantom{ }\hphantom{ }\hphantom{ }\hphantom{ }\textsf{TwicePlusOne}~\texttt{\symbol{40}}\textsf{f}~\textsf{elt}~\texttt{\symbol{44}}~\textsf{map}~\textsf{f}\texttt{\symbol{39}}~\textsf{lst}\texttt{\symbol{41}}\\[0.5\baselineskip]
    \textbf{\textsf{end}}}
  \end{displaymath}
  \par
  \subsection{Alternation}
  \par
  In Section~\ref{latex_lib_label_1}, labels are separated from trees dynamically. The \textsf{pass} function verifies that trees and labels are interleaved properly, and fails if they are not.
  \par
  In this section, instead, we consider a variant of ${\upalpha}$~\textsf{PowerList}\texttt{\symbol{46}}\textsf{t} where every odd position contains a tree, and every even position contains an element. More generally, we define a type \texttt{\symbol{40}}${\upomega}$\texttt{\symbol{44}}${\upeta}$\texttt{\symbol{41}}~\textsf{AlternatingPowerList}\texttt{\symbol{46}}\textsf{t} where odd positions have type ${\upomega}$, and even positions have type ${\upeta}$. Such a list should have the following pattern:
  \begin{itemize}
    \item \texttt{\symbol{91}}${\upomega}$\texttt{\symbol{59}\symbol{40}}${\upeta}$\texttt{\symbol{44}}${\upomega}$\texttt{\symbol{41}\symbol{59}\symbol{40}\symbol{40}}${\upeta}$\texttt{\symbol{44}}${\upomega}$\texttt{\symbol{41}\symbol{44}\symbol{40}}${\upeta}$\texttt{\symbol{44}}${\upomega}$\texttt{\symbol{41}\symbol{41}\symbol{93}}
  \end{itemize}
  After the first element, which must have type ${\upomega}$, there is no difference between even and odd positions: indeed, excluding the first element, we are actually building an \texttt{\symbol{40}}${\upeta}$\texttt{\symbol{42}}${\upomega}$\texttt{\symbol{41}}~\textsf{PowerList}\texttt{\symbol{46}}\textsf{t}. Hence the definition:
  \begin{displaymath}
    \parbox{0.9\linewidth}{\textbf{\textsf{module}}~\textsf{AlternatingPowerList}~\texttt{\symbol{61}}~\textbf{\textsf{struct}}\\[0.5\baselineskip]
    \hphantom{ }\hphantom{ }\textbf{\textsf{type}}~\texttt{\symbol{40}}${\upomega}$\texttt{\symbol{44}}${\upeta}$\texttt{\symbol{41}}~\textsf{t}~\texttt{\symbol{61}}\\
    \hphantom{ }\hphantom{ }\hphantom{ }\hphantom{ }\hspace*{0.1em}\rule[-0.6ex]{1.sp}{1.\baselineskip}\hspace*{0.2em}~\textsf{Zero}\\
    \hphantom{ }\hphantom{ }\hphantom{ }\hphantom{ }\hspace*{0.1em}\rule[-0.6ex]{1.sp}{1.\baselineskip}\hspace*{0.2em}~\textsf{TwicePlusOne}~\textsf{of}~${\upomega}$~\texttt{\symbol{42}}~\texttt{\symbol{40}}${\upeta}$\texttt{\symbol{42}}${\upomega}$\texttt{\symbol{41}}~\textsf{PowerList}\texttt{\symbol{46}}\textsf{t}\\[0.5\baselineskip]
    \textbf{\textsf{end}}}
  \end{displaymath}
  \par
  For brevity, let us write \textsf{PL} and \textsf{APL} for \textsf{PowerList} and \textsf{AlternatingPowerList} respectively.
  \par
  Using these alternating power lists, we can define a version of the \textsf{pass} function free of \textsf{assert}~\textsf{false}. Indeed, consider an alternating power list of length at least $3$:  it is of the form \textsf{APL}\texttt{\symbol{46}}\textsf{TwicePlusOne}~\texttt{\symbol{40}}\textsf{a}\texttt{\symbol{44}}~\textsf{PL}\texttt{\symbol{46}}\textsf{TwicePlusOne}~\texttt{\symbol{40}\symbol{40}}\textsf{b}\texttt{\symbol{44}}\textsf{c}\texttt{\symbol{41}\symbol{44}}~\textsf{lst}\texttt{\symbol{41}\symbol{41}}, where \textsf{lst} has type \texttt{\symbol{40}\symbol{40}}${\upeta}$\texttt{\symbol{42}}${\upomega}$\texttt{\symbol{41}\symbol{42}\symbol{40}}${\upeta}$\texttt{\symbol{42}}${\upomega}$\texttt{\symbol{41}\symbol{41}}~\textsf{PowerList}\texttt{\symbol{46}}\textsf{t}. The \textsf{pass} function of Section~\ref{latex_lib_label_1}, as it happens, manipulates its arguments by groups of four elements: basically, \textsf{pass} is simply a \textsf{map} over \textsf{lst}.
  \par
  We hence define the function \textsf{pass}
  which joins the trees in a list of length ${2}^{k+2}-1$\footnote{To ensure that its argument list has at least three elements, \textsf{pass} takes the three first elements as extra arguments. In other words \textsf{pass}~\textsf{t}~\texttt{\symbol{40}}\textsf{x}\texttt{\symbol{44}}\textsf{s}\texttt{\symbol{41}}~\textsf{l} is meant to be read as \textsf{pass}~\texttt{\symbol{40}}\textsf{APL}\texttt{\symbol{46}}\textsf{TwicePlusOne}~\texttt{\symbol{40}}\textsf{t}~\texttt{\symbol{44}}~\textsf{PL}\texttt{\symbol{46}}\textsf{TwicePlusOne}~\texttt{\symbol{40}\symbol{40}}\textsf{x}\texttt{\symbol{44}}\textsf{s}\texttt{\symbol{41}\symbol{44}}\textsf{l}\texttt{\symbol{41}\symbol{41}\symbol{41}}.}, producing a list of length ${2}^{k+1}-1$. The function \textsf{loop} is virtually unchanged from Section~\ref{latex_lib_label_1}, except it acts on power lists:
  \begin{displaymath}
    \parbox{0.9\linewidth}{\textbf{\textsf{let}}~\textsf{pass}~\textsf{left}~\texttt{\symbol{40}}\textsf{root}\texttt{\symbol{44}}\textsf{right}\texttt{\symbol{41}}~\textsf{apl}~\texttt{\symbol{61}}\\
    \hphantom{ }\hphantom{ }\textbf{\textsf{let}}~\textsf{join\texttt{\symbol{95}}up}~\texttt{\symbol{40}\symbol{40}}\textsf{single}\texttt{\symbol{44}}\textsf{left}\texttt{\symbol{41}\symbol{44}\symbol{40}}\textsf{root}\texttt{\symbol{44}}\textsf{right}\texttt{\symbol{41}\symbol{41}}~\texttt{\symbol{61}}\\
    \hphantom{ }\hphantom{ }\hphantom{ }\hphantom{ }\textsf{single}\texttt{\symbol{44}}~\textsf{Node}~\texttt{\symbol{40}}\textsf{left}\texttt{\symbol{44}}\textsf{root}\texttt{\symbol{44}}\textsf{right}\texttt{\symbol{41}}\\
    \hphantom{ }\hphantom{ }\textbf{\textsf{in}}\\
    \hphantom{ }\hphantom{ }\textsf{APL}\texttt{\symbol{46}}\textsf{TwicePlusOne}~\texttt{\symbol{40}}~\textsf{Node}~\texttt{\symbol{40}}\textsf{left}\texttt{\symbol{44}}\textsf{root}\texttt{\symbol{44}}\textsf{right}\texttt{\symbol{41}}~\texttt{\symbol{44}}~\textsf{PL}\texttt{\symbol{46}}\textsf{map}~\textsf{join\texttt{\symbol{95}}up}~\textsf{apl}\texttt{\symbol{41}}\\[0.5\baselineskip]
    \textbf{\textsf{let}}~\textbf{\textsf{rec}}~\textsf{loop}~\texttt{\symbol{58}}~${\upepsilon}$\texttt{\symbol{46}}~\texttt{\symbol{40}}${\upepsilon}$~\textsf{tree}\texttt{\symbol{44}}${\upepsilon}$\texttt{\symbol{41}}~\textsf{APL}\texttt{\symbol{46}}\textsf{t}~${\rightarrow}$~${\upepsilon}$~\textsf{tree}~\texttt{\symbol{61}}~\textbf{\textsf{function}}\\
    \hphantom{ }\hphantom{ }\hspace*{0.1em}\rule[-0.6ex]{1.sp}{1.\baselineskip}\hspace*{0.2em}~\textsf{APL}\texttt{\symbol{46}}\textsf{Zero}~${\rightarrow}$~\textsf{Leaf}\\
    \hphantom{ }\hphantom{ }\hspace*{0.1em}\rule[-0.6ex]{1.sp}{1.\baselineskip}\hspace*{0.2em}~\textsf{APL}\texttt{\symbol{46}}\textsf{TwicePlusOne}~\texttt{\symbol{40}}\textsf{tree}\texttt{\symbol{44}}\textsf{PL}\texttt{\symbol{46}}\textsf{Zero}\texttt{\symbol{41}}~${\rightarrow}$~\textsf{tree}\\
    \hphantom{ }\hphantom{ }\hspace*{0.1em}\rule[-0.6ex]{1.sp}{1.\baselineskip}\hspace*{0.2em}~\textsf{APL}\texttt{\symbol{46}}\textsf{TwicePlusOne}~\texttt{\symbol{40}}\textsf{tree}\texttt{\symbol{44}}\textsf{PL}\texttt{\symbol{46}}\textsf{TwicePlusOne}~\texttt{\symbol{40}}\textsf{pair}\texttt{\symbol{44}}\textsf{apl}\texttt{\symbol{41}\symbol{41}}~${\rightarrow}$\\
    \hphantom{ }\hphantom{ }\hphantom{ }\hphantom{ }\hphantom{ }\hphantom{ }\textsf{loop}~\texttt{\symbol{40}}\textsf{pass}~\textsf{tree}~\textsf{pair}~\textsf{apl}\texttt{\symbol{41}}}
  \end{displaymath}
  \par
  \subsection{Padding}
  \par
  Now that there is no more \textsf{assert}~\textsf{false} in the code of \textsf{loop}, we need to change the \textsf{complete} function of Section~\ref{latex_lib_label_1} so that it returns an \texttt{\symbol{40}}${\upalpha}$~\textsf{tree}\texttt{\symbol{44}}${\upalpha}$\texttt{\symbol{41}}~\textsf{APL}\texttt{\symbol{46}}\textsf{t} rather than a list. The heart of this section is a function which turns an ${\upalpha}$~\textsf{list} into an \texttt{\symbol{40}}${\upalpha}$\texttt{\symbol{42}}${\upalpha}$~\textsf{tree}\texttt{\symbol{41}}~\textsf{PL}\texttt{\symbol{46}}\textsf{t}. The final function, which produces an \texttt{\symbol{40}}${\upalpha}$~\textsf{tree}\texttt{\symbol{44}}${\upalpha}$\texttt{\symbol{41}}~\textsf{APL}\texttt{\symbol{46}}\textsf{t} is a simple wrapper around the former.
  \par
  We want to turn a list \textsf{lst} of length $n+1$ into a pair of its first element, converted into a tree, plus a power list of length $2\times (2^k-1)\geqslant  n$ representing its tail \textsf{tail}. Each element of the power list is a pair, whose first term is an element, and its second term is a tree of height at most one. In particular, the length of the returned power list is always even, so if \textsf{tail} has odd length, we will need to insert at least a \textsf{Leaf}. This suggests that we may inspect the parity of the length of \textsf{tail}, and insert an extra element precisely when it is odd. This leads to a slightly different padding procedure than that of Section~\ref{latex_lib_label_1}, in particular the leaves are not inserted at the same position, but it is inconsequential.
  \par
  An ${\upalpha}$~\textsf{list} of even length can be turned into an \texttt{\symbol{40}}${\upalpha}$\texttt{\symbol{42}}${\upalpha}$\texttt{\symbol{41}}~\textsf{list} whose length is halved. This turns out to be interesting for our recursion, since it mimics the inductive step of power lists. Also, in the case of even length, we need to distinguish the empty case from the non-empty case: the former will be turned into the empty power list \textsf{APL}\texttt{\symbol{46}}\textsf{Zero} while the latter will be turned into a power list of the form \textsf{APL}\texttt{\symbol{46}}\textsf{TwicePlusOne}\texttt{\symbol{40}\symbol{40}}\textsf{x}\texttt{\symbol{44}}\textsf{y}\texttt{\symbol{41}\symbol{44}}\textsf{l}\texttt{\symbol{41}}, where \textsf{x} and \textsf{y} correspond to the two first elements of \textsf{tail}. These different cases are represented in the following view:
  \begin{displaymath}
    \parbox{0.9\linewidth}{\textbf{\textsf{type}}~${\upalpha}$~\textsf{parity}~\texttt{\symbol{61}}\\
    \hphantom{ }\hphantom{ }\hspace*{0.1em}\rule[-0.6ex]{1.sp}{1.\baselineskip}\hspace*{0.2em}~\textsf{Empty}\\
    \hphantom{ }\hphantom{ }\hspace*{0.1em}\rule[-0.6ex]{1.sp}{1.\baselineskip}\hspace*{0.2em}~\textsf{Odd}~\textsf{of}~${\upalpha}$~\texttt{\symbol{42}}~\texttt{\symbol{40}}${\upalpha}$\texttt{\symbol{42}}${\upalpha}$\texttt{\symbol{41}}~\textsf{list}\\
    \hphantom{ }\hphantom{ }\hspace*{0.1em}\rule[-0.6ex]{1.sp}{1.\baselineskip}\hspace*{0.2em}~\textsf{Even}~\textsf{of}~\texttt{\symbol{40}}${\upalpha}$\texttt{\symbol{42}}${\upalpha}$\texttt{\symbol{41}}~\texttt{\symbol{42}}~\texttt{\symbol{40}}${\upalpha}$\texttt{\symbol{42}}${\upalpha}$\texttt{\symbol{41}}~\textsf{list}\\[0.5\baselineskip]
    \textbf{\textsf{let}}~\textsf{pair\texttt{\symbol{95}}up}~\textsf{lst}~\texttt{\symbol{61}}\\
    \hphantom{ }\hphantom{ }\textbf{\textsf{let}}~\textsf{succ}~\textsf{elt}~\texttt{\symbol{61}}~\textbf{\textsf{function}}\\
    \hphantom{ }\hphantom{ }\hphantom{ }\hphantom{ }\hspace*{0.1em}\rule[-0.6ex]{1.sp}{1.\baselineskip}\hspace*{0.2em}~\textsf{Empty}~${\rightarrow}$~\textsf{Odd}~\texttt{\symbol{40}}\textsf{elt}\texttt{\symbol{44}}~\texttt{\symbol{91}\symbol{93}\symbol{41}}\\
    \hphantom{ }\hphantom{ }\hphantom{ }\hphantom{ }\hspace*{0.1em}\rule[-0.6ex]{1.sp}{1.\baselineskip}\hspace*{0.2em}~\textsf{Odd}~\texttt{\symbol{40}}\textsf{b}\texttt{\symbol{44}}\textsf{pairs}\texttt{\symbol{41}}~${\rightarrow}$~\textsf{Even}~\texttt{\symbol{40}\symbol{40}}\textsf{elt}\texttt{\symbol{44}}\textsf{b}\texttt{\symbol{41}\symbol{44}}~\textsf{pairs}\texttt{\symbol{41}}\\
    \hphantom{ }\hphantom{ }\hphantom{ }\hphantom{ }\hspace*{0.1em}\rule[-0.6ex]{1.sp}{1.\baselineskip}\hspace*{0.2em}~\textsf{Even}~\texttt{\symbol{40}}\textsf{bc}\texttt{\symbol{44}}\textsf{pairs}\texttt{\symbol{41}}~${\rightarrow}$~\textsf{Odd}~\texttt{\symbol{40}}\textsf{elt}\texttt{\symbol{44}}~\textsf{bc}\texttt{:\hspace*{-0.25em}:}\textsf{pairs}\texttt{\symbol{41}}\\
    \hphantom{ }\hphantom{ }\textbf{\textsf{in}}\\
    \hphantom{ }\hphantom{ }\textsf{List}\texttt{\symbol{46}}\textsf{fold\texttt{\symbol{95}}right}~\textsf{succ}~\textsf{lst}~\textsf{Empty}}
  \end{displaymath}
  \par
  The padding function itself, \textsf{of\texttt{\symbol{95}}list}, is at first sight far from intuitive. Let us recall that we want to turn a list of labels of arbitrary length, into a power list of pairs. A label can be thought of as a bit of weight $2^0$, while a pair of labels would be a bit of weight $2^1$, and so on. At first, all our bits have weight $2^0$ and consists in one label each. We can build bits of higher weight by pairing up two bits of the same weight. A bit made up only of labels is called \emph{pure}. We can also double the weight of a bit by interlacing leaves with it (with the function \textsf{pad}), but this gives a bit made of pairs of labels and trees, call them \emph{impure}. Lastly, we can also transmute a pure bit into an impure bit of the same weight (with the function \textsf{coerce}), by replacing odd-position labels by trees of height one.
  \par
  Each recursive step consists in taking a list of pure bits of the same weight $2^k$, and outputing exactly one impure bit of size ${2}^{k+1}$, plus a list of pure bits of weight ${2}^{k+1}$, which is converted recursively. We thus obtain, successively, one bit of each weight from $2^1$ to $2^l$, for some $l$, encoding a list of length ${2}^{l+1} - 2$, as expected.
  \par
  At any recursive step, suppose first that the number of bits of weight $2^k$ is odd. As we need to compute only bits of weight ${2}^{k+1}$, one of them impure, we are forced to use \textsf{pad} on one bit, and to pair up the others. Suppose now that the number of bits of weight $2^k$ is even. In that case, we can pair them all into bits of weight ${2}^{k+1}$, and then use \textsf{coerce} on one of them to make the impure bit.
  \par
  The last difficulty is that \textsf{pad} and \textsf{coerce} both depend on the current weight of the bits, hence we need to update them at each recursive step. \textsf{pad} must add leaves between every two consecutive labels, in even positions, while \textsf{coerce} must upgrade every even-position label into a tree of height one. This leads to the following definition:
  \begin{displaymath}
    \parbox{0.9\linewidth}{\textbf{\textsf{module}}~\textsf{PowerList}~\texttt{\symbol{61}}~\textbf{\textsf{struct}}\\
    \hphantom{ }\hphantom{ }\hspace{1.em}\raisebox{-1.ex}[0.75\baselineskip][0.75\baselineskip]{{\vdots}}\\
    \hphantom{ }\hphantom{ }\textbf{\textsf{let}}~\textbf{\textsf{rec}}~\textsf{of\texttt{\symbol{95}}list}~\texttt{\symbol{58}}~${\upalpha}$~${\upbeta}$\texttt{\symbol{46}}~\texttt{\symbol{40}}${\upalpha}$${\rightarrow}$${\upbeta}$\texttt{\symbol{41}}~${\rightarrow}$~\texttt{\symbol{40}}${\upalpha}$\texttt{\symbol{42}}${\upalpha}$${\rightarrow}$${\upbeta}$\texttt{\symbol{41}}~${\rightarrow}$~${\upalpha}$~\textsf{list}~${\rightarrow}$~${\upbeta}$~\textsf{t}~\texttt{\symbol{61}}\\
    \hphantom{ }\hphantom{ }\hphantom{ }\hphantom{ }\textbf{\textsf{fun}}~\textsf{pad}~\textsf{coerce}~\textsf{bits}~${\rightarrow}$\\
    \hphantom{ }\hphantom{ }\hphantom{ }\hphantom{ }\hphantom{ }\hphantom{ }\textbf{\textsf{let}}~\textsf{pad}\texttt{\symbol{39}}~\texttt{\symbol{40}}\textsf{x}\texttt{\symbol{44}}\textsf{y}\texttt{\symbol{41}}~\texttt{\symbol{61}}~\texttt{\symbol{40}}\textsf{pad}~\textsf{x}\texttt{\symbol{44}}~\textsf{pad}~\textsf{y}\texttt{\symbol{41}}~\textbf{\textsf{in}}\\
    \hphantom{ }\hphantom{ }\hphantom{ }\hphantom{ }\hphantom{ }\hphantom{ }\textbf{\textsf{let}}~\textsf{coerce}\texttt{\symbol{39}}~\texttt{\symbol{40}}\textsf{x}\texttt{\symbol{44}}\textsf{y}\texttt{\symbol{41}}~\texttt{\symbol{61}}~\texttt{\symbol{40}}\textsf{coerce}~\textsf{x}\texttt{\symbol{44}}~\textsf{coerce}~\textsf{y}\texttt{\symbol{41}}~\textbf{\textsf{in}}\\
    \hphantom{ }\hphantom{ }\hphantom{ }\hphantom{ }\hphantom{ }\hphantom{ }\textbf{\textsf{match}}~\textsf{pair\texttt{\symbol{95}}up}~\textsf{bits}~\textbf{\textsf{with}}\\
    \hphantom{ }\hphantom{ }\hphantom{ }\hphantom{ }\hphantom{ }\hphantom{ }\hspace*{0.1em}\rule[-0.6ex]{1.sp}{1.\baselineskip}\hspace*{0.2em}~\textsf{Empty}~${\rightarrow}$~\textsf{Zero}\\
    \hphantom{ }\hphantom{ }\hphantom{ }\hphantom{ }\hphantom{ }\hphantom{ }\hspace*{0.1em}\rule[-0.6ex]{1.sp}{1.\baselineskip}\hspace*{0.2em}~\textsf{Odd}~\texttt{\symbol{40}}\textsf{a}\texttt{\symbol{44}}\textsf{pures}\texttt{\symbol{41}}~${\rightarrow}$~\textsf{TwicePlusOne}~\texttt{\symbol{40}}\textsf{pad}~\textsf{a}\texttt{\symbol{44}}~\textsf{of\texttt{\symbol{95}}list}~\textsf{pad}\texttt{\symbol{39}}~\textsf{coerce}\texttt{\symbol{39}}~\textsf{pures}\texttt{\symbol{41}}\\
    \hphantom{ }\hphantom{ }\hphantom{ }\hphantom{ }\hphantom{ }\hphantom{ }\hspace*{0.1em}\rule[-0.6ex]{1.sp}{1.\baselineskip}\hspace*{0.2em}~\textsf{Even}~\texttt{\symbol{40}}\textsf{ab}\texttt{\symbol{44}}\textsf{pures}\texttt{\symbol{41}}~${\rightarrow}$\\
    \hphantom{ }\hphantom{ }\hphantom{ }\hphantom{ }\hphantom{ }\hphantom{ }\hphantom{ }\hphantom{ }\hphantom{ }\hphantom{ }\textsf{TwicePlusOne}~\texttt{\symbol{40}}\textsf{coerce}~\textsf{ab}\texttt{\symbol{44}}~\textsf{of\texttt{\symbol{95}}list}~\textsf{pad}\texttt{\symbol{39}}~\textsf{coerce}\texttt{\symbol{39}}~\textsf{pures}\texttt{\symbol{41}}\\[0.5\baselineskip]
    \hphantom{ }\hphantom{ }\textbf{\textsf{end}}}
  \end{displaymath}
  \par
  With that function, we can conclude our implementation. Again writing \textsf{PL} and \textsf{APL} for \textsf{PowerList} and \textsf{AlternatingPowerList} respectively:
  \par
  \begin{displaymath}
    \parbox{0.9\linewidth}{\textbf{\textsf{module}}~\textsf{AlternatingPowerList}~\texttt{\symbol{61}}~\textbf{\textsf{struct}}\\
    \hphantom{ }\hphantom{ }\hspace{1.em}\raisebox{-1.ex}[0.75\baselineskip][0.75\baselineskip]{{\vdots}}\\
    \hphantom{ }\hphantom{ }\textbf{\textsf{let}}~\textsf{of\texttt{\symbol{95}}list}~\textsf{leaf}~\textsf{up}~\textsf{id}~\texttt{\symbol{61}}~\textbf{\textsf{function}}\\
    \hphantom{ }\hphantom{ }\hphantom{ }\hphantom{ }\hspace*{0.1em}\rule[-0.6ex]{1.sp}{1.\baselineskip}\hspace*{0.2em}~\texttt{\symbol{91}\symbol{93}}~${\rightarrow}$~\textsf{Zero}\\
    \hphantom{ }\hphantom{ }\hphantom{ }\hphantom{ }\hspace*{0.1em}\rule[-0.6ex]{1.sp}{1.\baselineskip}\hspace*{0.2em}~\textsf{a}\texttt{:\hspace*{-0.25em}:}\textsf{l}~${\rightarrow}$\\
    \hphantom{ }\hphantom{ }\hphantom{ }\hphantom{ }\hphantom{ }\hphantom{ }\hphantom{ }\hphantom{ }\textbf{\textsf{let}}~\textsf{pad}~\textsf{x}~\texttt{\symbol{61}}~\textsf{id}~\textsf{x}~\texttt{\symbol{44}}~\textsf{leaf}~\textbf{\textsf{in}}\\
    \hphantom{ }\hphantom{ }\hphantom{ }\hphantom{ }\hphantom{ }\hphantom{ }\hphantom{ }\hphantom{ }\textbf{\textsf{let}}~\textsf{coerce}~\texttt{\symbol{40}}\textsf{x}\texttt{\symbol{44}}\textsf{y}\texttt{\symbol{41}}~\texttt{\symbol{61}}~\textsf{id}~\textsf{x}~\texttt{\symbol{44}}~\textsf{up}~\textsf{y}~\textbf{\textsf{in}}\\
    \hphantom{ }\hphantom{ }\hphantom{ }\hphantom{ }\hphantom{ }\hphantom{ }\hphantom{ }\hphantom{ }\textsf{TwicePlusOne}~\texttt{\symbol{40}}\textsf{up}~\textsf{a}\texttt{\symbol{44}}~\textsf{PowerList}\texttt{\symbol{46}}\textsf{of\texttt{\symbol{95}}list}~\textsf{pad}~\textsf{coerce}~\textsf{l}\texttt{\symbol{41}}\\[0.5\baselineskip]
    \textbf{\textsf{end}}\\[0.5\baselineskip]
    \textbf{\textsf{let}}~\textsf{singleton}~\textsf{x}~\texttt{\symbol{61}}~\textsf{Node}\texttt{\symbol{40}}\textsf{Leaf}\texttt{\symbol{44}}\textsf{x}\texttt{\symbol{44}}\textsf{Leaf}\texttt{\symbol{41}}\\
    \textbf{\textsf{let}}~\textsf{balance}~\textsf{l}~\texttt{\symbol{61}}\\
    \hphantom{ }\hphantom{ }\textsf{loop}~\texttt{\symbol{40}}\textsf{APL}\texttt{\symbol{46}}\textsf{of\texttt{\symbol{95}}list}~\textsf{Leaf}~\textsf{singleton}~\texttt{\symbol{40}}\textbf{\textsf{fun}}~\textsf{e}${\rightarrow}$\textsf{e}\texttt{\symbol{41}}~\textsf{l}\texttt{\symbol{41}}}
  \end{displaymath}
  The final function, \textsf{balance}\texttt{\symbol{58}}${\upalpha}$~\textsf{list}~${\rightarrow}$~${\upalpha}$~\textsf{tree}, implements the same algorithm as Section~\ref{latex_lib_label_1} without any partial functions.
  \par
  What we may have lost in this section, compared to the simple algorithm, is the simplicity of the complexity analysis of the algorithm. The subtleties of the main functions require a finer analysis. Consider first the function \textsf{PL}\texttt{\symbol{46}}\textsf{map}: clearly the number of recursive calls depends only logarithmically on the number of elements in the power list. But each recursive call uses as its first argument a function twice as complex than the previous one. This leads to the following inequation over the complexity $C(n,m,f)$ of \textsf{map}, where $n$ is the number of elements in the power list, $m$ is the size of the elements in the power list, and $f$ is the complexity of the mapped function:
  \begin{displaymath}
    C(n,m,f) \leq  f(m) + C\left( \frac{n-1}{2}, 2m, k \mapsto  2 \times  f(k/2) + O(1)\right)  + O(1)
  \end{displaymath}
  From there, it is easy to prove that $C(n,m,f) = n.f(m) + O(n)$, so that \textsf{PL}\texttt{\symbol{46}}\textsf{map} runs indeed in linear-time, and so is \textsf{loop}. Similarly the complexity of \textsf{PL}\texttt{\symbol{46}}\textsf{of\texttt{\symbol{95}}list} can be described by a higher-order recursive inequation (almost the same as above, except that the complexity depends on two functions and the constant term is replaced by a linear term), whose solution gives also a linear-time complexity.
  \section{Turning to Coq}\label{latex_lib_label_3}
  \par
  There is still a property of the algorithm that the implementation of Section~\ref{latex_lib_label_2} does not make obvious: that the algorithm actually does build \emph{full} trees. In this section we shall build into the type of \textsf{balance} that its output is indeed full.
  \par
  To that effect, we will use Coq rather than Ocaml. Even if it is possible, with some effort, to represent full trees and implement the algorithm in Ocaml -- and relatively easy in Haskell -- a Coq implementation also gives us termination by construction. Coq forces every recursion to be structural, which will prove to be rather entertaining.
  \par
  At a superficial level, a visible difference with the Ocaml implementation is that \textsf{Powerlist}\texttt{\symbol{46}}\textsf{t} and \textsf{AlternatingPowerList}\texttt{\symbol{46}}\textsf{t} must be decorated with the $k$ such that the length is $2^k-1$: it is the structural recursion parameter of the \textsf{balance\_powerlist} function. Because it makes the code simpler, we will use a recursive definition rather than an inductive one:
  \begin{displaymath}
    \parbox{0.9\linewidth}{\textbf{\textsf{Module}}~\textsf{PowerList}\symbol{46}\\[0.5\baselineskip]
    \hphantom{ }\hphantom{ }\textbf{\textsf{Fixpoint}}~\textsf{T}~\symbol{40}\textsf{A}\symbol{58}\textbf{\textsf{Type}}\symbol{41}~\symbol{40}\textsf{k}\symbol{58}\textsf{nat}\symbol{41}~$\coloneqq $\\
    \hphantom{ }\hphantom{ }\hphantom{ }\hphantom{ }\textbf{\textsf{match}}~\textsf{k}~\textbf{\textsf{with}}\\
    \hphantom{ }\hphantom{ }\hphantom{ }\hphantom{ }\hspace{0.1em}\rule[-0.6ex]{1.sp}{1.\baselineskip}~0~${\Rightarrow}$~\textsf{unit}\symbol{58}\textbf{\textsf{Type}}\\
    \hphantom{ }\hphantom{ }\hphantom{ }\hphantom{ }\hspace{0.1em}\rule[-0.6ex]{1.sp}{1.\baselineskip}~\textsf{S}~\textsf{k}\symbol{39}~${\Rightarrow}$~\textsf{A}~$*$~\textsf{T}~\symbol{40}\textsf{A}$*$\textsf{A}\symbol{41}~\textsf{k}\symbol{39}\\
    \hphantom{ }\hphantom{ }\hphantom{ }\hphantom{ }\textbf{\textsf{end}}\symbol{46}\\[0.5\baselineskip]
    \textbf{\textsf{End}}~\textsf{PowerList}\symbol{46}}
  \end{displaymath}
  We will also need a version where \textsf{k} can be arbitrary. For that purpose we use Coq's type of dependent pairs \{~\textsf{n}\symbol{58}\textsf{nat}~\textsf{\&}~\textsf{F}~\textsf{n}\}. The constructor for dependent pairs is written ${\langle}$~\textsf{n}~\symbol{44}~\textsf{x}~${\rangle}$. The implicit version comes with constructors -- \textsf{tpo} stands for ``twice plus one'':
  \begin{displaymath}
    \parbox{0.9\linewidth}{\textbf{\textsf{Module}}~\textsf{PowerList}\symbol{46}\\
    \hphantom{ }\hphantom{ }\hspace{1.em}\raisebox{-1.ex}[0.75\baselineskip][0.75\baselineskip]{{\vdots}}\\
    \hphantom{ }\hphantom{ }\textbf{\textsf{Definition}}~\textsf{U}~\symbol{40}\textsf{A}\symbol{58}\textbf{\textsf{Type}}\symbol{41}~$\coloneqq $~\{~\textsf{k}\symbol{58}\textsf{nat}~\textsf{\&}~\textsf{T}~\textsf{A}~\textsf{k}~\}\symbol{46}\\
    \hphantom{ }\hphantom{ }\textbf{\textsf{Definition}}~\textsf{zero}~\{\textsf{A}\symbol{58}\textbf{\textsf{Type}}\}~\symbol{58}~\textsf{U}~\textsf{A}~$\coloneqq $~${\langle}$~0~\symbol{44}~\textsf{tt}~${\rangle}$\symbol{46}\\
    \hphantom{ }\hphantom{ }\textbf{\textsf{Definition}}~\textsf{tpo}~\{\textsf{A}\symbol{58}\textbf{\textsf{Type}}\}~\symbol{40}\textsf{a}\symbol{58}\textsf{A}\symbol{41}~\symbol{40}\textsf{l}\symbol{58}\textsf{U}~\symbol{40}\textsf{A}$*$\textsf{A}\symbol{41}\symbol{41}~\symbol{58}~\textsf{U}~\textsf{A}~$\coloneqq $\\
    \hphantom{ }\hphantom{ }\hphantom{ }\hphantom{ }\textsf{let}~\symbol{39}${\langle}$\textsf{k}\symbol{44}\textsf{l}${\rangle}$~$\coloneqq $~\textsf{l}~\textsf{in}\\
    \hphantom{ }\hphantom{ }\hphantom{ }\hphantom{ }${\langle}$~\textsf{S}~\textsf{k}~\symbol{44}~\symbol{40}\textsf{a}\symbol{44}\textsf{l}\symbol{41}~${\rangle}$\symbol{46}\\[0.5\baselineskip]
    \textbf{\textsf{End}}~\textsf{PowerList}}
  \end{displaymath}
  The definition of \textsf{AlternatingPowerList}\symbol{46}\textsf{T} and \textsf{AlternatingPowerList}\symbol{46}\textsf{U} are similar.
  \par
  \subsection{Full trees}
  \par
  To code full trees, we index trees by their height, and specify that leaves can happen only at height $0$ or $1$:
  \begin{displaymath}
    \parbox{0.9\linewidth}{\textbf{\textsf{Inductive}}~\textsf{FullTree}~\symbol{40}\textsf{A}\symbol{58}\textbf{\textsf{Type}}\symbol{41}~\symbol{58}~\textsf{nat}~${\rightarrow}$~\textbf{\textsf{Type}}~$\coloneqq $\\
    \hphantom{ }\hphantom{ }\hspace{0.1em}\rule[-0.6ex]{1.sp}{1.\baselineskip}~\textsf{Leaf}$_0$~\symbol{58}~\textsf{FullTree}~\textsf{A}~0\\
    \hphantom{ }\hphantom{ }\hspace{0.1em}\rule[-0.6ex]{1.sp}{1.\baselineskip}~\textsf{Leaf}$_1$~\symbol{58}~\textsf{FullTree}~\textsf{A}~1\\
    \hphantom{ }\hphantom{ }\hspace{0.1em}\rule[-0.6ex]{1.sp}{1.\baselineskip}~\textsf{Node}~\{\textsf{k}\symbol{58}\textsf{nat}\}~\symbol{58}~\textsf{FullTree}~\textsf{A}~\textsf{k}~${\rightarrow}$~\textsf{A}~${\rightarrow}$~\textsf{FullTree}~\textsf{A}~\textsf{k}~${\rightarrow}$~\textsf{FullTree}~\textsf{A}~\symbol{40}\textsf{S}~\textsf{k}\symbol{41}\symbol{46}}
  \end{displaymath}
  If we omitted the constructor \textsf{Leaf}$_1$, we would have a definition of complete binary trees: both subtrees of a node are complete binary trees of the same height. We allow the full trees to be incomplete by letting \textsf{Leaf}$_1$ take the place of nodes on the last level.
  \par
  Using the type \textsf{FullTree}~\textsf{A}~\textsf{k} in place of the type ${\upalpha}$~\textsf{tree}, the functions \textsf{pass} and \textsf{balance\_powerlist} are virtually unmodified\footnote{In fact, as can be seen from its type, \textsf{loop} only handles non-empty alternating power lists. This is due to a small technicality: the recursive step of \textsf{loop} is the case \textsf{S}~\symbol{40}\textsf{S}~\textsf{k}\symbol{41}, but Coq does not recognise \textsf{S}~\textsf{k} as a structural subterm of \textsf{S}~\symbol{40}\textsf{S}~\textsf{k}\symbol{41}, so the definition from Section~\ref{latex_lib_label_2} does not verifies Coq's structural recursion criterion. As a workaround, the empty case is moved to the \textsf{balance} function.} with respect to Section~\ref{latex_lib_label_2}. Only their types change to reflect the extra information:
  \begin{displaymath}
    \parbox{0.9\linewidth}{\textbf{\textsf{Definition}}~\textsf{pass}~\{\textsf{A}~\textsf{k}~\textsf{p}\}~\symbol{58}~\textsf{APL}\symbol{46}\textsf{T}~\symbol{40}\textsf{FullTree}~\textsf{A}~\symbol{40}\textsf{S}~\textsf{p}\symbol{41}\symbol{41}~\textsf{A}~\symbol{40}\textsf{S}~\symbol{40}\textsf{S}~\textsf{k}\symbol{41}\symbol{41}~${\rightarrow}$\\
    \hphantom{ }\hphantom{ }\hphantom{ }\hphantom{ }\hphantom{ }\textsf{APL}\symbol{46}\textsf{T}~\symbol{40}\textsf{FullTree}~\textsf{A}~\symbol{40}\textsf{S}~\symbol{40}\textsf{S}~\textsf{p}\symbol{41}\symbol{41}\symbol{41}~\textsf{A}~\symbol{40}\textsf{S}~\textsf{k}\symbol{41}\symbol{46}\\
    \textbf{\textsf{Fixpoint}}~\textsf{loop}~\{\textsf{A}~\textsf{k}~\textsf{p}\}~\symbol{58}~\textsf{APL}\symbol{46}\textsf{T}~\symbol{40}\textsf{FullTree}~\textsf{A}~\symbol{40}\textsf{S}~\textsf{p}\symbol{41}\symbol{41}~\textsf{A}~\symbol{40}\textsf{S}~\textsf{k}\symbol{41}~${\rightarrow}$\\
    \hphantom{ }\hphantom{ }\hphantom{ }\hphantom{ }\hphantom{ }\textsf{FullTree}~\textsf{A}~\symbol{40}\textsf{plus}~\textsf{k}~\symbol{40}\textsf{S}~\textsf{p}\symbol{41}\symbol{41}~\{\textsf{struct}~\textsf{k}\}\symbol{46}}
  \end{displaymath}
  The algorithm indeed builds only full trees.
  \par
  \subsection{Structural initialisation}
  \par
  The padding conversion from lists to power lists, in Section~\ref{latex_lib_label_2}, is not structural due to the use of \textsf{pair\texttt{\symbol{95}}up} in the recursive call. To tackle this recursion, we shall make use of another intermediate structure. What we need, essentially, is that all the calls to \textsf{pair\texttt{\symbol{95}}up} are pre-calculated, so the intermediate structure will be like \textsf{parity} except that the calls to \texttt{\symbol{40}}${\upalpha}$\texttt{\symbol{42}}${\upalpha}$\texttt{\symbol{41}}~\textsf{list} are replaced by inductive calls.
  \par
  As it turns out, this is another non-uniform datatype which corresponds to a numerical representation. Indeed, any natural number can be written in binary with digits $1$ and $2$ (but not $0$). In this system, for example, $8=1\times 2^2+1\times 2^1+2\times 2^0$ is represented as $112$. Here is the definition, where \textsf{tpo} reads ``twice plus one'' and \textsf{tpt} ``twice plus two'':
  \begin{displaymath}
    \parbox{0.9\linewidth}{\textbf{\textsf{Module}}~\textsf{BinaryList}\symbol{46}\\[0.5\baselineskip]
    \hphantom{ }\hphantom{ }\textbf{\textsf{Inductive}}~\textsf{T}~\symbol{40}\textsf{A}\symbol{58}\textbf{\textsf{Type}}\symbol{41}~\symbol{58}~\textbf{\textsf{Type}}~$\coloneqq $\\
    \hphantom{ }\hphantom{ }\hspace{0.1em}\rule[-0.6ex]{1.sp}{1.\baselineskip}~\textsf{zero}\\
    \hphantom{ }\hphantom{ }\hspace{0.1em}\rule[-0.6ex]{1.sp}{1.\baselineskip}~\textsf{tpo}~\symbol{40}\textsf{a}\symbol{58}\textsf{A}\symbol{41}~\symbol{40}\textsf{l}\symbol{58}\textsf{T}~\symbol{40}\textsf{A}$*$\textsf{A}\symbol{41}\symbol{41}\\
    \hphantom{ }\hphantom{ }\hspace{0.1em}\rule[-0.6ex]{1.sp}{1.\baselineskip}~\textsf{tpt}~\symbol{40}\textsf{a}~\textsf{b}\symbol{58}~\textsf{A}\symbol{41}~\symbol{40}\textsf{l}\symbol{58}\textsf{T}~\symbol{40}\textsf{A}$*$\textsf{A}\symbol{41}\symbol{41}\symbol{46}\\[0.5\baselineskip]
    \textbf{\textsf{End}}~\textsf{BinaryList}\symbol{46}}
  \end{displaymath}
  \par
  To turn a non-empty list into a \textsf{BinaryList}\symbol{46}\textsf{T}, all we need is a function \textsf{cons} of type \textsf{A}~${\rightarrow}$~\textsf{T}~\textsf{A}~${\rightarrow}$~\textsf{T}~\textsf{A} to add an element in front of the list. On the numerical representation side, it corresponds to adding $1$. It behaves like adding $1$ in the usual binary representation, except that $1$-s are turned into $2$-s without a carry and $2$-s into $1$-s while producing a carry:
  \begin{displaymath}
    \parbox{0.9\linewidth}{\textbf{\textsf{Module}}~\textsf{BinaryList}\symbol{46}\\
    \hphantom{ }\hphantom{ }\hspace{1.em}\raisebox{-1.ex}[0.75\baselineskip][0.75\baselineskip]{{\vdots}}\\
    \hphantom{ }\hphantom{ }\textbf{\textsf{Fixpoint}}~\textsf{cons}~\{\textsf{A}\}~\symbol{40}\textsf{a}\symbol{58}\textsf{A}\symbol{41}~\symbol{40}\textsf{l}\symbol{58}\textsf{T}~\textsf{A}\symbol{41}~\symbol{58}~\textsf{T}~\textsf{A}~$\coloneqq $\\
    \hphantom{ }\hphantom{ }\hphantom{ }\hphantom{ }\textbf{\textsf{match}}~\textsf{l}~\textbf{\textsf{with}}\\
    \hphantom{ }\hphantom{ }\hphantom{ }\hphantom{ }\hspace{0.1em}\rule[-0.6ex]{1.sp}{1.\baselineskip}~\textsf{zero}~${\Rightarrow}$~\textsf{tpo}~\textsf{a}~\textsf{zero}\\
    \hphantom{ }\hphantom{ }\hphantom{ }\hphantom{ }\hspace{0.1em}\rule[-0.6ex]{1.sp}{1.\baselineskip}~\textsf{tpo}~\textsf{b}~\textsf{l}~${\Rightarrow}$~\textsf{tpt}~\textsf{a}~\textsf{b}~\textsf{l}\\
    \hphantom{ }\hphantom{ }\hphantom{ }\hphantom{ }\hspace{0.1em}\rule[-0.6ex]{1.sp}{1.\baselineskip}~\textsf{tpt}~\textsf{b}~\textsf{c}~\textsf{l}~${\Rightarrow}$~\textsf{tpo}~\textsf{a}~\symbol{40}\textsf{cons}~\symbol{40}\textsf{b}\symbol{44}\textsf{c}\symbol{41}~\textsf{l}\symbol{41}\\
    \hphantom{ }\hphantom{ }\hphantom{ }\hphantom{ }\textbf{\textsf{end}}\symbol{46}\\[0.5\baselineskip]
    \hphantom{ }\hphantom{ }\textbf{\textsf{Definition}}~\textsf{of\_list}~\{\textsf{A}\}~\symbol{40}\textsf{l}\symbol{58}\textsf{list}~\textsf{A}\symbol{41}~\symbol{58}~\textsf{T}~\textsf{A}~$\coloneqq $\\
    \hphantom{ }\hphantom{ }\hphantom{ }\hphantom{ }\textsf{List}\symbol{46}\textsf{fold\_right}~\textsf{cons}~\textsf{zero}~\textsf{l}\symbol{46}\\[0.5\baselineskip]
    \textbf{\textsf{End}}~\textsf{BinaryList}\symbol{46}}
  \end{displaymath}
  Note that while \textsf{cons} takes, in the worst case, logarithmic time with respect to the length of the list, building a list by repeatedly using \textsf{cons} is still linear. Indeed, as previously mentioned, \textsf{cons} mimics the successor algorithm for binary numbers, whose amortized complexity is well-known to be constant.
  \par
  We also need a function which turns a \textsf{T}~\symbol{40}\textsf{A}$*$\textsf{A}\symbol{41} into a \textsf{T}~\textsf{A}. This is effectively multiplication by $2$. The lack of $0$ among the digits\footnote{The constructor \textsf{zero} represents an empty list of digits.} makes this process recursive. A simple presentation of the doubling algorithm consists in adding a $0$ at the end of the number, then eliminating the $0$ using the following equalities:
  \begin{displaymath}
    \left\{ 
    \begin{array}{clc}
      0 & = & \cdot \\
      x20 & = & x12\\
      x10 & = & x02\\
    \end{array}
    \right. 
  \end{displaymath}
  In terms of binary lists:
  \begin{displaymath}
    \parbox{0.9\linewidth}{\textbf{\textsf{Module}}~\textsf{BinaryList}\symbol{46}\\
    \hphantom{ }\hphantom{ }\hspace{1.em}\raisebox{-1.ex}[0.75\baselineskip][0.75\baselineskip]{{\vdots}}\\
    \hphantom{ }\hphantom{ }\textbf{\textsf{Fixpoint}}~\textsf{twice}~\{\textsf{A}\}~\symbol{40}\textsf{l}\symbol{58}\textsf{T}~\symbol{40}\textsf{A}$*$\textsf{A}\symbol{41}\symbol{41}~\symbol{58}~\textsf{T}~\textsf{A}~$\coloneqq $\\
    \hphantom{ }\hphantom{ }\hphantom{ }\hphantom{ }\textbf{\textsf{match}}~\textsf{l}~\textbf{\textsf{with}}\\
    \hphantom{ }\hphantom{ }\hphantom{ }\hphantom{ }\hspace{0.1em}\rule[-0.6ex]{1.sp}{1.\baselineskip}~\textsf{zero}~${\Rightarrow}$~\textsf{zero}\\
    \hphantom{ }\hphantom{ }\hphantom{ }\hphantom{ }\hspace{0.1em}\rule[-0.6ex]{1.sp}{1.\baselineskip}~\textsf{tpo}~\symbol{40}\textsf{a}\symbol{44}\textsf{b}\symbol{41}~\textsf{l}~${\Rightarrow}$~\textsf{tpt}~\textsf{a}~\textsf{b}~\symbol{40}\textsf{twice}~\textsf{l}\symbol{41}\\
    \hphantom{ }\hphantom{ }\hphantom{ }\hphantom{ }\hspace{0.1em}\rule[-0.6ex]{1.sp}{1.\baselineskip}~\textsf{tpt}~\symbol{40}\textsf{a}\symbol{44}\textsf{b}\symbol{41}~\textsf{cd}~\textsf{l}~${\Rightarrow}$~\textsf{tpt}~\textsf{a}~\textsf{b}~\symbol{40}\textsf{tpo}~\textsf{cd}~\textsf{l}\symbol{41}\\
    \hphantom{ }\hphantom{ }\hphantom{ }\hphantom{ }\textbf{\textsf{end}}\symbol{46}\\[0.5\baselineskip]
    \textbf{\textsf{End}}~\textsf{BinaryList}\symbol{46}}
  \end{displaymath}
  \par
  We can now write a structurally recursive padding function, using binary lists as the structural argument. As we do not know in advance the length of the produced list, a \textsf{PowerList}\symbol{46}\textsf{U} is returned. We write \textsf{BL} as a shorthand for \textsf{BinaryList}:
  \begin{displaymath}
    \parbox{0.9\linewidth}{\textbf{\textsf{Module}}~\textsf{PowerList}\symbol{46}\\
    \hphantom{ }\hphantom{ }\hspace{1.em}\raisebox{-1.ex}[0.75\baselineskip][0.75\baselineskip]{{\vdots}}\\
    \hphantom{ }\hphantom{ }\textbf{\textsf{Fixpoint}}~\textsf{of\_binary\_list}~\{\textsf{A}~\textsf{X}\}~\symbol{40}\textsf{d}\symbol{58}\textsf{A}${\rightarrow}$\textsf{X}\symbol{41}~\symbol{40}\textsf{f}\symbol{58}\textsf{A}$*$\textsf{A}${\rightarrow}$\textsf{X}\symbol{41}~\symbol{40}\textsf{l}\symbol{58}\textsf{BL}\symbol{46}\textsf{T}~\textsf{A}\symbol{41}~\symbol{58}~\textsf{U}~\textsf{X}~$\coloneqq $\\
    \hphantom{ }\hphantom{ }\hphantom{ }\hphantom{ }\textbf{\textsf{match}}~\textsf{l}~\textbf{\textsf{with}}\\
    \hphantom{ }\hphantom{ }\hphantom{ }\hphantom{ }\hspace{0.1em}\rule[-0.6ex]{1.sp}{1.\baselineskip}~\textsf{BL}\symbol{46}\textsf{zero}~${\Rightarrow}$~\textsf{zero}\\
    \hphantom{ }\hphantom{ }\hphantom{ }\hphantom{ }\hspace{0.1em}\rule[-0.6ex]{1.sp}{1.\baselineskip}~\textsf{BL}\symbol{46}\textsf{tpo}~\textsf{a}~\textsf{l}~${\Rightarrow}$\\
    \hphantom{ }\hphantom{ }\hphantom{ }\hphantom{ }\hphantom{ }\hphantom{ }\textsf{tpo}~\symbol{40}\textsf{d}~\textsf{a}\symbol{41}~\symbol{40}\textsf{of\_binary\_list}~\symbol{40}\textsf{d}${\times}$\textsf{d}\symbol{41}~\symbol{40}\textsf{f}${\times}$\textsf{f}\symbol{41}~\textsf{l}\symbol{41}\\
    \hphantom{ }\hphantom{ }\hphantom{ }\hphantom{ }\hspace{0.1em}\rule[-0.6ex]{1.sp}{1.\baselineskip}~\textsf{BL}\symbol{46}\textsf{tpt}~\textsf{a}~\textsf{b}~\textsf{l}~${\Rightarrow}$\\
    \hphantom{ }\hphantom{ }\hphantom{ }\hphantom{ }\hphantom{ }\hphantom{ }\textsf{tpo}~\symbol{40}\textsf{f}~\symbol{40}\textsf{a}\symbol{44}\textsf{b}\symbol{41}\symbol{41}~\symbol{40}\textsf{of\_binary\_list}~\symbol{40}\textsf{d}${\times}$\textsf{d}\symbol{41}~\symbol{40}\textsf{f}${\times}$\textsf{f}\symbol{41}~\textsf{l}\symbol{41}\\
    \hphantom{ }\hphantom{ }\hphantom{ }\hphantom{ }\textbf{\textsf{end}}\symbol{46}\\[0.5\baselineskip]
    \textbf{\textsf{End}}~\textsf{PowerList}\symbol{46}}
  \end{displaymath}
  Where \textsf{g}${\times}$\textsf{f} is the function which maps \symbol{40}\textsf{x}\symbol{44}\textsf{y}\symbol{41} to \symbol{40}\textsf{g}~\textsf{x}\symbol{44}\textsf{f}~\textsf{y}\symbol{41}.
  \par
  The rest follows straightforwardly, and we can define the following functions which conclude the algorithm (\textsf{BL}, \textsf{PL}, and \textsf{APL} stand for \textsf{BinaryList}, \textsf{PowerList}, and \textsf{AlternatingPowerList} respectively):
  \begin{displaymath}
    \parbox{0.9\linewidth}{\textbf{\textsf{Module}}~\textsf{AlternatingPowerList}\symbol{46}\\
    \hphantom{ }\hphantom{ }\hspace{1.em}\raisebox{-1.ex}[0.75\baselineskip][0.75\baselineskip]{{\vdots}}\\
    \hphantom{ }\hphantom{ }\textbf{\textsf{Definition}}~\textsf{of\_binary\_list}~\{\textsf{A}~\textsf{Odd}~\textsf{Even}\}\\
    \hphantom{ }\hphantom{ }\hphantom{ }\hphantom{ }\hphantom{ }\hphantom{ }\symbol{40}\textsf{d}\symbol{58}\textsf{Odd}\symbol{41}~\symbol{40}\textsf{f}\symbol{58}\textsf{A}${\rightarrow}$\textsf{Odd}\symbol{41}~\symbol{40}\textsf{g}\symbol{58}\textsf{A}${\rightarrow}$\textsf{Even}\symbol{41}~\symbol{40}\textsf{l}\symbol{58}\textsf{BL}\symbol{46}\textsf{T}~\textsf{A}\symbol{41}~\symbol{58}~\textsf{U}~\textsf{Odd}~\textsf{Even}~$\coloneqq $\\
    \hphantom{ }\hphantom{ }\hphantom{ }\hphantom{ }\textbf{\textsf{match}}~\textsf{l}~\textbf{\textsf{with}}\\
    \hphantom{ }\hphantom{ }\hphantom{ }\hphantom{ }\hspace{0.1em}\rule[-0.6ex]{1.sp}{1.\baselineskip}~\textsf{BL}\symbol{46}\textsf{zero}~${\Rightarrow}$~\textsf{zero}\\
    \hphantom{ }\hphantom{ }\hphantom{ }\hphantom{ }\hspace{0.1em}\rule[-0.6ex]{1.sp}{1.\baselineskip}~\textsf{BL}\symbol{46}\textsf{tpo}~\textsf{a}~\textsf{l}~${\Rightarrow}$\\
    \hphantom{ }\hphantom{ }\hphantom{ }\hphantom{ }\hphantom{ }\hphantom{ }\textsf{let}~\textsf{d}\symbol{39}~\textsf{x}~$\coloneqq $~\symbol{40}~\textsf{g}~\textsf{x}~\symbol{44}~\textsf{d}~\symbol{41}~\textsf{in}\\
    \hphantom{ }\hphantom{ }\hphantom{ }\hphantom{ }\hphantom{ }\hphantom{ }\textsf{tpo}~\symbol{40}\textsf{f}~\textsf{a}\symbol{41}~\symbol{40}\textsf{PL}\symbol{46}\textsf{of\_binary\_list}~\textsf{d}\symbol{39}~\symbol{40}\textsf{g}${\times}$\textsf{f}\symbol{41}~\symbol{40}\textsf{BL}\symbol{46}\textsf{twice}~\textsf{l}\symbol{41}\symbol{41}\\
    \hphantom{ }\hphantom{ }\hphantom{ }\hphantom{ }\hspace{0.1em}\rule[-0.6ex]{1.sp}{1.\baselineskip}~\textsf{BL}\symbol{46}\textsf{tpt}~\textsf{a}~\textsf{b}~\textsf{l}~${\Rightarrow}$\\
    \hphantom{ }\hphantom{ }\hphantom{ }\hphantom{ }\hphantom{ }\hphantom{ }\textsf{let}~\textsf{d}\symbol{39}~\textsf{x}~$\coloneqq $~\symbol{40}~\textsf{g}~\textsf{x}~\symbol{44}~\textsf{d}~\symbol{41}~\textsf{in}\\
    \hphantom{ }\hphantom{ }\hphantom{ }\hphantom{ }\hphantom{ }\hphantom{ }\textsf{tpo}~\symbol{40}\textsf{f}~\textsf{a}\symbol{41}~\symbol{40}\textsf{PL}\symbol{46}\textsf{of\_binary\_list}~\textsf{d}\symbol{39}~\symbol{40}\textsf{g}${\times}$\textsf{f}\symbol{41}~\symbol{40}\textsf{BL}\symbol{46}\textsf{tpo}~\textsf{b}~\textsf{l}\symbol{41}\symbol{41}\\
    \hphantom{ }\hphantom{ }\hphantom{ }\hphantom{ }\textbf{\textsf{end}}\symbol{46}\\[0.5\baselineskip]
    \hphantom{ }\hphantom{ }\textbf{\textsf{Definition}}~\textsf{of\_list}~\{\textsf{A}~\textsf{Odd}~\textsf{Even}\}\\
    \hphantom{ }\hphantom{ }\hphantom{ }\hphantom{ }\hphantom{ }\hphantom{ }\symbol{40}\textsf{d}\symbol{58}\textsf{Odd}\symbol{41}~\symbol{40}\textsf{f}\symbol{58}\textsf{A}${\rightarrow}$\textsf{Odd}\symbol{41}~\symbol{40}\textsf{g}\symbol{58}\textsf{A}${\rightarrow}$\textsf{Even}\symbol{41}~\symbol{40}\textsf{l}\symbol{58}\textsf{list}~\textsf{A}\symbol{41}~\symbol{58}~\textsf{U}~\textsf{Odd}~\textsf{Even}~$\coloneqq $\\
    \hphantom{ }\hphantom{ }\hphantom{ }\hphantom{ }\textsf{of\_binary\_list}~\textsf{d}~\textsf{f}~\textsf{g}~\symbol{40}\textsf{BL}\symbol{46}\textsf{of\_list}~\textsf{l}\symbol{41}\symbol{46}\\[0.5\baselineskip]
    \textbf{\textsf{End}}~\textsf{AlternatingPowerList}\symbol{46}\\[0.5\baselineskip]
    \textbf{\textsf{Definition}}~\textsf{singleton}~\{\textsf{A}\symbol{58}\textbf{\textsf{Type}}\}~\symbol{40}\textsf{x}\symbol{58}\textsf{A}\symbol{41}~\symbol{58}~\textsf{FullTree}~\textsf{A}~1~$\coloneqq $\\
    \hphantom{ }\hphantom{ }\textsf{Node}~\textsf{Leaf}$_0$~\textsf{x}~\textsf{Leaf}$_0$\symbol{46}\\[0.5\baselineskip]
    \textbf{\textsf{Definition}}~\textsf{balance}~\{\textsf{A}\symbol{58}\textbf{\textsf{Type}}\}~\symbol{40}\textsf{l}\symbol{58}\textsf{list}~\textsf{A}\symbol{41}~\symbol{58}~\{~\textsf{k}\symbol{58}\textsf{nat}~\textsf{\&}~\textsf{FullTree}~\textsf{A}~\textsf{k}~\}~$\coloneqq $\\
    \hphantom{ }\hphantom{ }\textsf{let}~\symbol{39}${\langle}$\textsf{k}\symbol{44}\textsf{l}${\rangle}$~$\coloneqq $~\textsf{APL}\symbol{46}\textsf{of\_list}~\textsf{Leaf}$_1$~\textsf{singleton}~\symbol{40}\textbf{\textsf{fun}}~\textsf{x}${\Rightarrow}$\textsf{x}\symbol{41}~\textsf{l}~\textsf{in}\\
    \hphantom{ }\hphantom{ }\textbf{\textsf{match}}~\textsf{k}~\textbf{\textsf{with}}\\
    \hphantom{ }\hphantom{ }\hspace{0.1em}\rule[-0.6ex]{1.sp}{1.\baselineskip}~0~${\Rightarrow}$~\textbf{\textsf{fun}}~$\_$~${\Rightarrow}$~${\langle}$~0~\symbol{44}~\textsf{Leaf}$_0$~${\rangle}$\\
    \hphantom{ }\hphantom{ }\hspace{0.1em}\rule[-0.6ex]{1.sp}{1.\baselineskip}~\textsf{S}~\textsf{k}~${\Rightarrow}$~\textbf{\textsf{fun}}~\textsf{l}~${\Rightarrow}$~${\langle}$~\textsf{plus}~\textsf{k}~1~\symbol{44}~\textsf{loop}~\textsf{l}~${\rangle}$\\
    \hphantom{ }\hphantom{ }\textbf{\textsf{end}}~\textsf{l}\symbol{46}}
  \end{displaymath}
  \section{Conclusion}
  \par
  The \textsf{balance} function of Section~\ref{latex_lib_label_3} is, by virtue of its type alone, a total function which turns lists into full binary trees. Yet, to the cost of using intermediary data-structures, it effectively implements the algorithm of Section~\ref{latex_lib_label_1}.
  \par
  The missing piece is to prove that the infix traversal of \textsf{balance}~\textsf{l} is indeed \textsf{l}. The infix traversal of a (full) tree is represented in Coq with the functions
  \begin{displaymath}
    \parbox{0.9\linewidth}{\textbf{\textsf{Fixpoint}}~\textsf{list\_of\_full\_tree\_n}~\{\textsf{A}~\textsf{n}\}~\symbol{40}\textsf{t}\symbol{58}\textsf{FullTree}~\textsf{A}~\textsf{n}\symbol{41}~\symbol{58}~\textsf{list}~\textsf{A}~$\coloneqq $\\
    \hphantom{ }\hphantom{ }\textbf{\textsf{match}}~\textsf{t}~\textbf{\textsf{with}}\\
    \hphantom{ }\hphantom{ }\hspace{0.1em}\rule[-0.6ex]{1.sp}{1.\baselineskip}~\textsf{Leaf}$_0$~${\Rightarrow}$~\symbol{91}\symbol{93}\\
    \hphantom{ }\hphantom{ }\hspace{0.1em}\rule[-0.6ex]{1.sp}{1.\baselineskip}~\textsf{Leaf}$_1$~${\Rightarrow}$~\symbol{91}\symbol{93}\\
    \hphantom{ }\hphantom{ }\hspace{0.1em}\rule[-0.6ex]{1.sp}{1.\baselineskip}~\textsf{Node}~$\_$~\textsf{t}$_1$~\textsf{x}~\textsf{t}$_2$~${\Rightarrow}$\\
    \hphantom{ }\hphantom{ }\hphantom{ }\hphantom{ }\hphantom{ }\textsf{list\_of\_full\_tree\_n}~\textsf{t}$_1$~$+\!\!\!+$~\symbol{91}\textsf{x}\symbol{93}~$+\!\!\!+$~\textsf{list\_of\_full\_tree\_n}~\textsf{t}$_2$\\
    \hphantom{ }\hphantom{ }\textbf{\textsf{end}}\symbol{46}\\[0.5\baselineskip]
    \textbf{\textsf{Definition}}~\textsf{list\_of\_full\_tree}~\{\textsf{A}\}~\symbol{40}\textsf{t}\symbol{58}\{~\textsf{k}\symbol{58}\textsf{nat}~\textsf{\&}~\textsf{FullTree}~\textsf{A}~\textsf{k}~\}\symbol{41}~\symbol{58}~\textsf{list}~\textsf{A}~$\coloneqq $\\
    \hphantom{ }\hphantom{ }\textsf{list\_of\_full\_tree\_n}~\symbol{40}\textsf{projT2}~\textsf{t}\symbol{41}\symbol{46}}
  \end{displaymath}
  We can then state the theorem:
  \begin{displaymath}
    \parbox{0.9\linewidth}{\textbf{\textsf{Theorem}}~\textsf{balance\_preserves\_order}~\textsf{A}~\symbol{40}\textsf{l}\symbol{58}\textsf{list}~\textsf{A}\symbol{41}~\symbol{58}\\
    \hphantom{ }\hphantom{ }\hphantom{ }\hphantom{ }\textsf{list\_of\_full\_tree}~\symbol{40}\textsf{balance}~\textsf{l}\symbol{41}~\symbol{61}~\textsf{l}\symbol{46}}
  \end{displaymath}
  \par
  The proof is short and straightforward: we define a traversal function for each intermediate structure; and state a variant of \textsf{balance\_preserves\_order} for each intermediate function. Proving the intermediate lemmas is not difficult and can be mostly automatised: we use a very simple generic automated tactic, which discharges most goals. This theorem concludes our easy formal proof of the balancing algorithm.
  \bibliography{library}
\end{document}